\documentclass[aps,preprint,epsfig,rotate]{revtex4}
\begin{document}
\title{Hyperfine structure splitting of the bound $S(L = 0)-$states
       in the symmetric muonic molecular ions}

 \author{Alexei M. Frolov}
 \email[E--mail address: ]{afrolov@uwo.ca}

\affiliation{Department of Chemistry\\
 University of Western Ontario, London, Ontario N6H 5B7, Canada}

\date{\today}

\begin{abstract}

The hyperfine structure splittings are determined for all five bound $S(L = 
0)-$states in the three symmetric muonic molecular ions: $pp\mu, dd\mu$ 
and $tt\mu$. The expectation values of all interparticle delta-functions
used in our calculations have been determined in recent highly accurate
computations.

\end{abstract}

\maketitle
\newpage

In our earlier study \cite{Fro01} we analyzed the hyperfine structure splitting 
of the bound $S(L = 0)-$states in the muonic molecular ions: $pd\mu, pt\mu$ and
$dt\mu$. The main goal of this work is to consider the hyperfine structure 
splitting in the symmetric muonic molecular ions $pp\mu, dd\mu$ and $tt\mu$.
Here and everywhere below in this study the notation $p$ designates the proton,
$d$ means deuteron (the nucleus of deuterium) and $t$ stands for the trition (or
the nucleus of tritium). As is well known there are five bound $S(L = 0)-$states 
in these (symmetric) muonic molecular ions. The ground states are stable in each 
of these ions, while the excited $S(L = 0)-$states are stable only in the heavy 
$dd\mu$ and $tt\mu$ ions. In general, the analysis of the hyperfine structure in 
the symmetric muonic molecular ions is slightly more complicated than analogous 
analysis for the non-symmetric ions. On the other hand, the hyperfine structures
of the symmetric muonic molecular ions are relatively simple and can be explained 
by using a few transparent physical ideas.

The general formula for the hyperfine structure splitting (or hyperfine 
splitting, for short) for an arbitrary (symmetric) three-body muonic molecular 
ion $aa\mu$ is written in the following form (in atomic units) (see, e.g., 
\cite{LLQ})
\begin{eqnarray}
 (\Delta H)_{h.s.} = \frac{2 \pi}{3} \alpha^2 \frac{g_a g_a}{m^2_p}
 \langle \delta({\bf r}_{aa}) \rangle ({\bf s}_a \cdot {\bf s}_a)+
 \frac{2 \pi}{3} \alpha^2 \frac{g_a g_{\mu}}{m_p m_{\mu}}
  \langle \delta({\bf r}_{a\mu}) \rangle ({\bf s}_a \cdot {\bf s}_{\mu}) 
 \nonumber \\
 + \frac{2 \pi}{3} \alpha^2 \frac{g_a g_{\mu}}{m_p m_{\mu}}
  \langle \delta({\bf r}_{a\mu}) \rangle ({\bf s}_a \cdot {\bf s}_{\mu})
 \label{e1}
\end{eqnarray}
where $\alpha = \frac{e^2}{\hbar c}$ is the fine structure constant, $m_{\mu}$ 
and $m_p$ are the muon and proton masses, respectively. The factors $g_{\mu}$
and $g_{a}$ are the corresponding $g-$factors. The expression, Eq.(\ref{e1}), for
$(\Delta H)_{h.s.}$ is, in fact, an operator in the total spin space which has 
the dimension $(2 s_a + 1)^2 (2 s_{\mu} + 1)$. Since the second and third terms 
in Eq.(1) are identical, then we can reduce Eq.(\ref{e1}) to the 
form
\begin{eqnarray}
 (\Delta H)_{h.s.} = \frac{2 \pi}{3} \alpha^2 \frac{g_a g_a}{m^2_p}
 \langle \delta({\bf r}_{aa}) \rangle ({\bf s}_a \cdot {\bf s}_a)+
 \frac{2 \pi}{3} \alpha^2 \frac{g_a g_{\mu}}{m_p m_{\mu}}
 \langle \delta({\bf r}_{a\mu}) \rangle ({\bf S}_{aa} \cdot 
 {\bf s}_{\mu}) \label{e2}
\end{eqnarray}
where ${\bf S}_{aa} = ({\bf s}_a + {\bf s}_a)$ is the total spin of the pair
of identical particles, i.e. the two nuclei of the hydrogen isotopes ($a = p, d$ 
and/or $t$).

The formula, Eq.(\ref{e2}), allows one to make a few qualitative predictions about 
the hyperfine structure of the symmetric muonic molecular ions. First, it is clear
that the classifications of the levels of hyperfine structure must be based on
the total spin of the two `symmetric' nuclei ${\bf S}_{aa}$. The absolute values of 
the spin ${\bf S}_{aa}$ are always non-negative integer numbers, i.e. $\mid {\bf 
S}_{aa} \mid = 0, 1, 2, \ldots$. For instance, in the case of two protons $p$ and/or 
two tritons $t$ one finds $\mid {\bf S}_{aa} \mid = 0, 1$, while for the two 
deuterons $\mid {\bf S}_{aa} \mid = 0, 1, 2$. The energy of the hyperfine state 
with ${\bf S}_{aa} = 0$ is determined only by the first term in Eq.(\ref{e2}). It is 
clear that such an energy is very small, since the expectation values $\langle 
\delta({\bf r}_{aa}) \rangle$ in all muonic molecular ions are very small. As follows 
from actual computations of muonic molecular ions all these values are less than $4 
\cdot 10^{-5}$ (in muon atomic units), while the expectation values of the muon-nuclear 
delta-functions are in $10^{4} - 10^{6}$ times larger. Briefly, we can say 
that the energy of this hyperfine state (with $J = \frac12$ for the $pp\mu$ and $tt\mu$ 
ions) is determined by the spin-spin interaction between the two heavy nuclei (muon's 
spin does not contribute). The overall contribution from the first term in Eq.(\ref{e2}) 
rapidly (exponentially) decreases when the mass of the heavy particle increases. 
Formally, the first term in Eq.(\ref{e2}) is very small already for the $pp\mu$ ion. 
However, for the $dd\mu$ and $tt\mu$ ions its contibution is negligible. This means 
that in the first approximation the hyperfine structure of the symmetric muonic 
molecular ions can be explained by using only one term for the muon-nuclear spin-spin 
interaction. This leads to some `additional' symmetry observed for the actual levels of 
hyperfine structure of heavy $dd\mu$ and $tt\mu$ ions (see below).   

As is well known the spin of the negatively charged muon $\mu^{-}$ equals $\frac12$ 
and the spins of the proton $p$ and triton $t$ also equal $\frac12$. Therefore, the 
hyperfine structure of the $pp\mu$ and $tt\mu$ ions must include eight levels which 
form three following groups: (1) the group of four spin states with $J = \frac32$, 
(2) the upper group of two states with $J = \frac12$ and (3) the lower group of two 
states with $J = \frac12$. The hyperfine energy of one of the two groups of states 
with $J = \frac12$ is very close to zero. The same classification of the hyperfine 
structure levels is true for the excited $S(L = 0)-$state in the $tt\mu$ ion. Here 
and everywhere below the notation $J$ stands for the total spin (or total momentum 
${\bf J} = {\bf L} + {\bf S} = {\bf S} = {\bf S}_{aa} + {\bf s}_{\mu} $, for the 
$S(L = 0)-$states) of the three-body ion.

The hyperfine structure of the $dd\mu$ ion is substantially different. In the $dd\mu$ 
ion one finds eighteen levels of hyperfine structure which are separated into five 
different groups: one group with $J = \frac52$ (six states), two different groups of 
states (upper and lower groups) with $J = \frac32$ (four states in each), two 
different groups of states (upper and lower groups) with $J = \frac12$ (two states 
in each). 

In our calculations we have used the following numerical values for the constants 
and factors in Eq.(\ref{e2}): $\alpha = 7.297352586 \cdot 10^{-3}, g_{\mu} = 
-2.0023218396$ \cite{CRC} and $m_p = 1836.152701 m_e, m_{\mu} = 206.768262 m_e$. The 
$g-$factors for the proton, deuteron and triton are deteremined from the formulas: 
$g_p = \frac{{\cal M}_d}{I_p}, g_d = \frac{{\cal M}_d}{I_d}$ and $g_t = \frac{{\cal 
M}_t}{I_t}$, where ${\cal M}_p = 2.792847386, {\cal M}_d = 0.857438230$ and 
${\cal M}_t = 2.97896247745$ are the magnetic moments (in nuclear magnetons) of the 
proton, deuteron and triton, respectively. Here the spins of the proton, deuteron 
and triton are designated in by the letter $I$ with the corresponding index: $I_p = 
\frac12, I_d = 1$ and $I_t = \frac12$. In Eqs.(\ref{e1}) - (\ref{e2}) these values 
are designated differently. In highly accurate computations of the expectation 
values of delta-functions we have used the following masses of the deuteron and 
triton: $m_d$ = 3680.483014 $m_e$ and $m_t$ = 5496.92158 $m_e$. These
masses are often used in modern highly accurate calculations of muonic molecular ions
(see, e.g., \cite{FrWa2011}). 

The convergence of the expectation values of the nuclear-nuclear (or $pp-$) and 
nuclear-muonic (or $p\mu-$) delta-functions is illustrated in Table I for the $pp\mu$ 
ion. The convergence of these expectation values  computed for other bound $S(L = 
0)-$states in the $dd\mu$ and $tt\mu$ ions is very similar to the results presented in 
Table I for the $pp\mu$ ion. The hyperfine structure and energy splittings between the 
corresponding levels for all five bound $S(L = 0)-$states in the three muonic molecular 
ions $pp\mu, dd\mu$ and $tt\mu$ can be found in Tables II and III. In atomic physics 
these values are traditionally given in $MHz$. The corresponding conversion factor is 
6.57968392061$\cdot 10^9$ $MHz/a.u.$ In Tables II and III the excited states are 
designated by the asterisk used as the upper index, e.g., $(dd\mu)^{*}$ and 
$(tt\mu)^{*}$. Such a system of notation is often used for muonic molecular ions. 

Tables II and III contain both the energies of the levels of hyperfine structure 
($\epsilon_{J}$) and hyperfine structure splitting ($\Delta(J_1 \rightarrow J_2)$). As 
we have predicted (see above) one of the hyperfine levels has a very small energy. As 
follows from Tables II and III this level corresponds to $J=\frac12$. In the $tt\mu$ 
ion the hyperfine energies of this state are $\approx$ 11.0591 $MHz$ and $\approx$ 
12.4307 $MHz$ for the ground and first excited states, respectively. In the $dd\mu$
ion the energies of the analogous levels are 6.8996 $MHz$ and 4.7378 $MHz$, 
respectively. Briefly, this means that the overall contribution of the nuclear-nuclear 
spin interaction is very small for the both $dd\mu$ and $tt\mu$ ions. This directly 
follows from the known fact (see, e.g., \cite{Fro99}) that the expectation values of 
nuclear-nuclear delta-functions in the $dd\mu$ and $tt\mu$ ions are very small. For 
instance, for the $dd\mu$ and $tt\mu$ ions the expectation values of nuclear-nuclear 
delta-functions are $\langle \delta_{dd} \rangle \approx 2.43871205 \cdot 10^{-6}$ 
($m.a.u.$), $\langle \delta_{dd} \rangle \approx 1.67460229 \cdot 10^{-6}$ ($m.a.u.$),
$\langle \delta_{tt} \rangle \approx 2.15893994 \cdot 10^{-7}$ ($m.a.u.$) and
$\langle \delta_{tt} \rangle \approx 2.42670033 \cdot 10^{-7}$ ($m.a.u.$), for the ground 
and excited states, respectively. Finally, the observed hyperfine structure of these two 
ions is mainly (99.9999 \%) related to the muon-nuclear spin-spin interactions. In the 
$pp\mu$ ion the situation is slightly different, but even for this ion the overall 
contribution of the muon-nuclear spin interaction(s) is substantially larger than the 
contribution from the nuclear-nuclear spin-spin interaction.

\newpage 

\begin{table}[tbp]
    \caption{The convergence of the $\langle \delta_{p\mu} \rangle$ and $\langle 
             \delta_{pp} \rangle$ expectation values for the ground (bound) $S(L 
             = 0)-$state of the $pp\mu$ molecular ion (in muon-atomic units).}
      \begin{center}
      \begin{tabular}{lll}
        \hline\hline
 $N$ & $\langle \delta_{31} \rangle$ & $\langle \delta_{21} \rangle$ \\  
      \hline
  3300 & 1.315008614364$\cdot 10^{-1}$ & 3.9370034861$\cdot 10^{-5}$ \\

  3500 & 1.315008614369$\cdot 10^{-1}$ & 3.9370034722$\cdot 10^{-5}$ \\

  3700 & 1.315008614374$\cdot 10^{-1}$ & 3.9370034782$\cdot 10^{-5}$ \\

  3840 & 1.315008614378$\cdot 10^{-1}$ & 3.9370034773$\cdot 10^{-5}$ \\
       \hline\hline
   \end{tabular}
   \end{center}
   \end{table}
\begin{table}[tbp]
    \caption{The hyperfine structure and hyperfine structure splitting of the 
             bound $S(L = 0)-$states of the $pp\mu$ and $tt\mu$ ions (in $MHz$).}
      \begin{center}
      \begin{tabular}{lllll}
        \hline\hline
      & $pp\mu$ & $tt\mu$ & $(tt\mu)^{*}$\\
      \hline
$\epsilon_{J=\frac32}$ &  1.256448515$\cdot 10^7$ &  1.736310113$\cdot 10^7$ &  1.510018118$\cdot 10^7$ \\

$\epsilon_{J=\frac12}$ &  1.772596177$\cdot 10^3$ &  1.105911127$\cdot 10^1$ &  1.243070661$\cdot 10^1$ \\

$\epsilon_{J=\frac12}$ & -2.513074289$\cdot 10^7$ & -3.472621366$\cdot 10^7$ & -3.020037480$\cdot 10^7$ \\
      \hline
$\Delta(\frac32 \rightarrow \frac12)$ & 1.256271251$\cdot 10^7$ & 1.735309024$\cdot 10^7$ & 1.510016875$\cdot 10^7$ \\

$\Delta(\frac12 \rightarrow \frac12)$ & 2.513251549$\cdot 10^7$ & 3.472622472$\cdot 10^7$ & 3.020038723$\cdot 10^7$ \\
       \hline\hline
   \end{tabular}
   \end{center}
   \end{table}
\begin{table}[tbp]
    \caption{The hyperfine structure and hyperfine structure splitting of the 
             bound $S(L = 0)-$states of the $dd\mu$ ion (in $MHz$).}
      \begin{center}
      \begin{tabular}{llll}
        \hline\hline
         & $dd\mu$ & $(dd\mu)^{*}$\\
      \hline
$\epsilon_{J=\frac52}$ &  4.656669271$\cdot 10^6$ &  4.023227167$\cdot 10^6$ \\

$\epsilon_{J=\frac32}$ &  2.328339810$\cdot 10^6$ &  2.011617137$\cdot 10^6$ \\

$\epsilon_{J=\frac12}$ &  6.899579465$\cdot 10^0$ &  4.737788941$\cdot 10^0$ \\

$\epsilon_{J=\frac12}$ & -4.656669271$\cdot 10^6$ & -4.023227167$\cdot 10^6$ \\

$\epsilon_{J=\frac32}$ & -6.985012530$\cdot 10^6$ & -6.034846672$\cdot 10^6$ \\
      \hline
       \hline
$\Delta(\frac52 \rightarrow \frac32)$ & 2.328329461$\cdot 10^6$ & 2.01161003$\cdot 10^6$ \\

$\Delta(\frac32 \rightarrow \frac12)$ & 2.328332911$\cdot 10^6$ & 2.01161239$\cdot 10^6$ \\

$\Delta(\frac12 \rightarrow \frac12)$ & 4.656676170$\cdot 10^6$ & 4.02323191$\cdot 10^6$ \\

$\Delta(\frac12 \rightarrow \frac32)$ & 2.328343259$\cdot 10^6$ & 2.01161951$\cdot 10^6$ \\
       \hline\hline
   \end{tabular}
   \end{center}
   \end{table}
\end{document}